\documentclass{article}

\usepackage{arxiv}

\usepackage[utf8]{inputenc} 
\usepackage[T1]{fontenc}    
\usepackage{hyperref}       
\usepackage{url}            
\usepackage{booktabs}       
\usepackage{amsfonts}       
\usepackage{nicefrac}       
\usepackage{microtype}      
\usepackage{lipsum}
\usepackage{graphicx}

\begin{document}
\vspace*{0.35in}
{\LARGE
\textbf{\newline{SCREEN\lowercase{et}: A Multi-view Deep Convolutional Neural Network} \vspace*{0.07in}\newline{for Classification of High-resolution Synthetic Mammographic} \vspace*{0.07in}\newline{Screening Scans}}
}
\vspace*{0.25in}
\newline
\\
\textbf{Saeed Seyyedi}\textsuperscript{1,2,*},
\textbf{Margaret J.~Wong}\textsuperscript{2},
\textbf{Debra M.~Ikeda}\textsuperscript{2},
\textbf{Curtis P.~Langlotz}\textsuperscript{1,2}
\\
\vspace*{0.08in}

1 Center for Artificial Intelligence in Medicine \& Imaging (AIMI), Stanford University, Stanford, CA

2 Department of Radiology, Stanford University, Stanford, CA
\vspace*{0.15in}
\\
$\*$ *seyyedi@stanford.edu

\vspace*{0.35in}

\begin{abstract}
\textbf{Background:} Digital breast tomosynthesis (DBT) has recently emerged as a promising modality to improve screening sensitivity and accuracy and showed significant reduction in the number of recalls comparing to digital mammography. Only a few recent studies have used small datasets of digital breast tomosynthesis cases to develop deep learning models for classification and detection of cancer.

\textbf{Purpose:} To develop and evaluate the accuracy of a multi-view deep learning approach to the analysis of high-resolution synthetic mammograms from digital breast tomosynthesis screening cases, and to assess the effect on accuracy of image resolution and training set size.

\textbf{Materials and Methods:} In a retrospective study, 21,264 screening digital breast tomosynthesis (DBT) exams obtained at our institution between 2012 and 2019 were collected along with associated radiology reports. The 2D synthetic mammographic images from these exams, with varying resolutions and data set sizes, were used to train a multi-view deep convolutional neural network (MV-CNN) to classify screening images into BI-RADS classes (BI-RADS 0, 1 and 2) before evaluation on a held-out set of exams.

\textbf{Results: }Area under the receiver operating characteristic curve (AUC) for BI-RADS 0 vs non-BI-RADS 0 class was 0.912 for the MV-CNN trained on the full dataset. The model obtained accuracy of 84.8\% , recall of 95.9\% and precision of 95.0\%. This AUC value decreased when the same model was trained with 50\% and 25\% of images (AUC = 0.877, P=0.010 and 0.834, P=0.009 respectively). Also, the performance dropped when the same model was trained using images that were under-sampled by 1/2 and 1/4 (AUC = 0.870, P=0.011 and 0.813, P=0.009 respectively). 

\textbf{Conclusion:} This deep learning model classified high-resolution synthetic mammography scans into normal vs needing further workup using tens of thousands of high-resolution images. Smaller training data sets and lower resolution images both caused significant decrease in performance.

\end{abstract}


\section{INTRODUCTION}
According to the World Health Organization, every year breast cancer kills more than 500,000 women around the world \cite{2014WHO}.  In the United States in 2020 there will be an estimated 276,480 new breast cancer cases with an estimated 42,170 breast cancer deaths. The five-year survival rate of women with advanced breast cancer is 28.1\% according to SEER 2010-2016 data. However, in settings where early screening and treatments are accessible, the five-year survival rate exceeds 80\% in the USA (localized disease 98.9\% and regional disease 85.7\% five-year survival) \cite{0Cancer}. 

In a standard breast cancer screening process, a digital mammogram or digital breast tomosynthesis study is obtained. Next, a radiologist reads the study and classifies the findings according to the Breast Imaging Reporting and Data System (BI-RADS) from the American College of Radiology \cite{EB2013ACR}. An abnormal finding typically requires that the patient be recalled to acquire additional mammographic views or imaging by other modalities, such as ultrasound or MRI, as part of an additional diagnostic workup. If a lesion remains equivocal or suspicious for cancer, further assessment is recommended with either short interval follow-up or tissue sampling. Although the introduction of screening mammography has contributed to an important reduction in breast cancer-related mortality \cite{Holford2015Contribution}, the majority of patients with positive screening studies ultimately do not have cancer. At the same time, reading and analysis of screening mammograms by highly trained specialists can be costly and time-consuming.  Interpretation is also prone to variation and error due to the subtle differences between lesions and background fibroglandular tissue, breast cancer morphology, differences in breast cancer pathologies, the nonrigid nature of the breast, and the relatively small proportion of cancers in a screening population of women at average risk \cite{Vlahiotis2018Analysis, Kooi2017Classifying}. Furthermore, waiting times associated with a patient recall after a positive screening may cause patient anxiety. 

Given that screening scans are interpreted by highly specialized radiologists, there are several advantages in automating this process. Both technologic and workflow solutions have been proposed to improve radiologist interpretive performance and efficiency. Computer-aided detection (CAD) models, which detect and mark suspicious findings on mammograms and aim to improve radiologists’ sensitivity ultimately may not have improved radiologists’ performance in sensitivity or specificity in real world clinical practice \cite{Tim2003Influence, Lehman2015Diagnostic}. 

However, the recent success of deep convolutional neural networks (CNNs) in computer vision tasks has resulted in an influx of research applying CNNs to medical imaging problems including digital mammography \cite{Carneiro2017Deep, Dhungel2015Automated, Ribli2018Detecting, Gastounioti2018Using, abraham2020machine, abdar2019cwv, janzen2019oa06}. A recent study showed that a deep learning model trained on a subset of 66,661 screening mammograms has the potential to reduce radiologist workload and significantly improve specificity without harming sensitivity \cite{Yala2019Deep}. In another study, a multi-view deep convolutional neural network was developed using a dataset of digital mammograms to perform classification based on BI-RADS classes \cite{Geras0High-Resolution}. 

Digital breast tomosynthesis (DBT) has recently emerged as a promising modality to improve screening sensitivity and accuracy and showed significant reduction in the number of recalls comparing to digital mammography \cite{Aase2019randomized}. Only a few recent studies have used small datasets of digital breast tomosynthesis cases to develop deep learning models for classification and detection of cancer. For example, Samala et al. \cite{Samala0Computer-aided} proposed a computer aided model for detection of microcalcification clusters using a dataset of 307 DBT volumes. In another work, Mendel et al. \cite{Mendel2019Transfer} compared the performance of computer-aided diagnosis on DBT images to that of conventional digital mammography. 

Our study is motivated by recent success of using DBT in breast cancer screening and successful applications of CNNs for medical image analysis and pattern recognition tasks. Unlike previous studies, we have developed a multi-view CNN (MV-CNN) as a classification model by using 82,943 high resolution synthetic mammographic images gathered from a total of 21,264 screening digital breast tomosynthesis examinations acquired at our organization from 2012 to 2019. Our deep learning model handles high-resolution images of multiple views without image down-sampling.  

Our contributions are as follows: (1) The proposed MV-CNN model classifies high-resolution synthetic screening mammograms from digital breast tomosynthesis without any shape hypothesis or user-interactive parameter settings, and it learns discriminative features automatically from a large amount of image data; (2), our data set of 82,943 synthetic mammography images is to our knowledge the largest dataset used for machine learning on this modality thus far, which enables us to assess the impact of training data set size; (3) The MV-CNN integrates multiple branches that can learn deep features from both MedioLateral Oblique (MLO) and Craniocaudal (CC) synthetic mammographic views; (4) we show that reducing training set size or image resolution both cause significant degradation in performance.

In the sections below, we first provide a detailed description of our MV-CNN.  Next, we introduce our experimental dataset and performance evaluation techniques. Then, we provide the quantitative performance results for our model and experiments. Finally, we discuss the implications of our results.

\section{MATERIALS AND METHODS}

Data acquisition and processing in this study were approved by our institutional review board and were compliant with the Health Insurance Portability and Accountability Act. Written informed consent was waived because of the retrospective nature of this study. All images and associated data were de-identified.

\subsection{Dataset}
This retrospective study employed 21,264 screening digital breast tomosynthesis examinations performed at our institute from January 1, 2012 to December 31, 2019, together with the associated text reports. The dataset included 82,943 synthetic right and left MLO and CC views of the breast.

We randomly split this dataset into three subsets to create a training subset which was used to train the neural network model, a validation subset to optimize the parameters, and a test set of images for final model evaluation. The test set of images was held out and not used to train or optimize the model. The distribution of training, validation and test dataset for different BI-RADS categories is shown in Table 1. Corresponding screening mammography structured reports were used to extract the labels for each image. Each report includes the information for BI-RADS scores where BI-RADS 0 indicates that further evaluation is required due to a suspicious abnormality such as a mass or calcification.

BI-RADS 1 means that the mammogram is negative, and no evident signs of cancer are found. Finally, BI-RADS 2 means that the mammogram is benign, with no apparent cancer; specifically, the mammogram is normal but other benign findings such as normal lymph nodes, typically benign calcifications or cysts are described in the report \cite{Bittner2010Guide}.

\begin{table}[]
\centering
\caption{distribution of imaging data with different BI-RADS used for training, validation and test in this study. (Format: Number of cases / Number of images)}
\label{tab:my-table}
\begin{tabular}{r|cccc}
 & \textbf{BI-RADS 0} & \textbf{BI-RADS 1} & \textbf{BI-RADS 2} &  \\ \cline{1-4}
 \textbf{Training} & 2,649 / 10,087 & 9,235 / 36,047 & 6,984 / 27,343 &  \\ \cline{1-4}
 \textbf{Validation} & 294 / 1,146 & 1,026 / 4,062 & 776 / 3,064  & \\ \cline{1-4}
 \textbf{Test} & 100 / 398 & 100 / 400 & 100 / 396& 
\end{tabular}
\end{table}

\subsection{Data Preprocessing and Labeling}
All images underwent histogram equalization for contrast enhancement \cite{DAVIES2012Computer}. Then, the images were normalized by computing the mean ($\mu$) and the standard deviation ($\sigma$) of its pixels and then subtracting $\mu$ from each pixel and dividing it by $\sigma$. Because of the large size of the data set and the lack of clinically realistic transformations, no data augmentations were used in our study. All the left breast images were flipped so the breast was always on the same side of image.
Because the original images varied in size and the large part of each image was empty background space, we cropped all images to the same size, $2200\times 1600$ pixels. 

Many image classification and object recognition studies of natural images down-sample the original high-resolution images to improve memory utilization and reduce the computational complexity of their models. For example, the best performing model for classification task of ImageNet challenge in 2015 was trained with images that were downscaled to $224\times 224$ pixels. Down-sampling works well for natural scenes, where the coarse features such as shapes, colors and other global descriptors are used to perform the classification and detection tasks. However, downscaling of mammographic images is not desirable due to the presence of finer features that may represent early cancer or ductal carcinoma in situ (DCIS). To preserve fine spatial features, we did not apply any down-sampling on the images prior to training for classification task. 

Image level labels were assigned by extracting the BI-RADS code for each breast from the radiology report generated at the time of the exam.

\subsection{Model Architecture}

Our multi-view CNN (MV-CNN) incorporates two parallel branches that process synthetic mammography images from MLO and CC views respectively. Each branch includes a modified ResNet-50 network to extract the view-specific features \cite{He2015Deep}. To decrease the computational cost of handling full-resolution images, we modified multiple convolution and pooling layers by setting the strides to be equal to two in the first two convolutional layers and by increasing the stride in first pooling layer, similar to the approach in [14]. Using this method, we reduced the size of feature maps in the network. We also averaged the features before the concatenation step to reduce the dimensionality of view-specific vectors. Using these two modifications, we were able to use the full-resolution synthetic mammography images as input to our network while the model remained computationally tractable.

The network was trained by stochastic gradient descent with backpropagation \cite{Ruineihart1986Learning}. The parameters of networks were initialized using the He initialization method \cite{He2015Delving} and were learned using the Adam loss function \cite{Kingma2014Adam} with initial learning rate of $10^{-5}$ and mini-batch size of 4. We trained the network for up to 100 epochs.

\begin{figure}
  \centering
  \includegraphics[width=12cm]{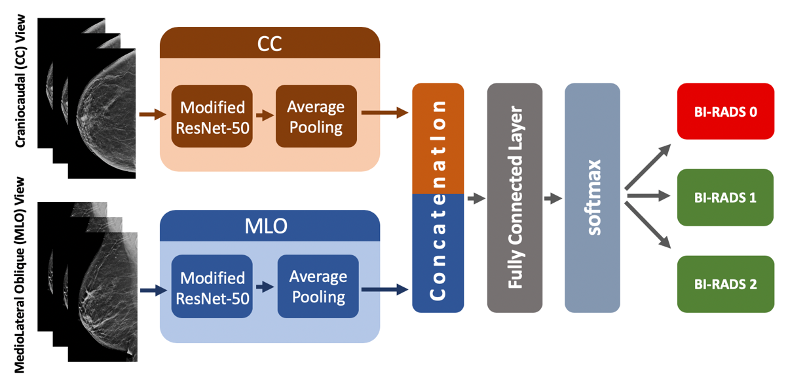}
  \caption{Illustration of Multi-view Convolutional Neural Network architecture for BI-RADs classification. This model contains two branches aiming at capturing features from CC and MLO image views. Each branch takes a high-resolution image as an input and includes a modified ResNet-50 network as a feature extractor and an average pooling step to reduce the feature map size. After these steps, view-specific feature maps are concatenated to form a vector which is fed to a fully connected layer followed by a softmax layer producing output distribution $p(y|x)$.}
  \label{fig:fig1}
\end{figure}

\subsection{Performance Evaluation}

Performance was assessed by using area under the receiver operating characteristic curve (AUC), accuracy, precision, and recall metrics. These measurements were computed and reported for the test set in each experiment. 

We computed three different ROC curves and AUC for the three BI-RADS classes (i.e., 0, 1, 2) using one vs all methodology and computed the macro average of three AUCs. Because the most clinically relevant distinction is differentiation between suspicious cases and non-suspicious ones, our reported results focus on ROC curves for the BI-RADS 0 class. 

The DeLong nonparametric statistical test was used to assess the statistical differences among AUC values, wherein P values less than 0.05 were considered to indicate a significant difference \cite{Delong2016Comparing}. 

\subsection{Effect of Dataset Size}

To explore the impact of dataset size on the performance of our model and to validate our earlier claim that we need a large dataset for our deep convolutional model to perform well, we trained different models on the randomly selected subsets of training data with 100\%, 50\% and 25\% of the original training set.

\subsection{Effect of Image Resolution}

To train our model using high resolution synthetic mammograms, we modified our network to accept the high-resolution images without any further down-sampling. To validate our approach, we studied the effect of image resolution reduction on model’s performance. We used the full dataset and scaled both dimensions of input images by the factor of 1/4 and 1/2. 

\subsection{Model Interpretation}

We used saliency maps to obtain a qualitative understanding of how the trained model arrives at its predictions. These maps were created by calculating the contribution of each pixel to the probability that the trained model assigns to the true BI-RADS class. This approach was performed by applying backpropagation through the trained model starting from the single probability value assigned to the true BI-RADS class back to the pixels of the original images that produced that probability.

\section{RESULTS}

We first measured model performance using the whole dataset for training. AUC, accuracy, recall, and precision metrics are shown in Table 2. Figure 2 shows the ROC curves for all BI-RADS classes in the left and average ROC curve in the right. Figure 3 illustrates the confusion matrix on the testing dataset. 

We then investigated the effect of changes in training set size and image resolution. 

\begin{table}[]
\centering
\caption{Performance Metrics for Different BI-RADS numbers}
\label{tab:my-table}
\begin{tabular}{r|c c c c}
 \textbf{Category} & \textbf{AUC} & \textbf{Accuracy (\%)} & \textbf{Recall (\%)} & \textbf{Precision (\%)} \\ \cline{1-5}
  \textbf{BI-RADS 0 vs. others} & 0.912 & 84.76 & 95.89 & 95.00  \\ \cline{1-5}
  \textbf{BI-RADS 1 vs. others} & 0.905 & 82.33 & 92.47 & 91.16  \\ \cline{1-5}
  \textbf{BI-RADS 2 vs. others} & 0.900 & 82.00 & 91.78 & 90.32  \\ \cline{1-5}
  \textbf{Mac Average} & 0.906 & 83.03 & 93.38 & 92.16
\end{tabular}
\end{table}

\begin{figure}
  \centering
  \includegraphics[width=14cm]{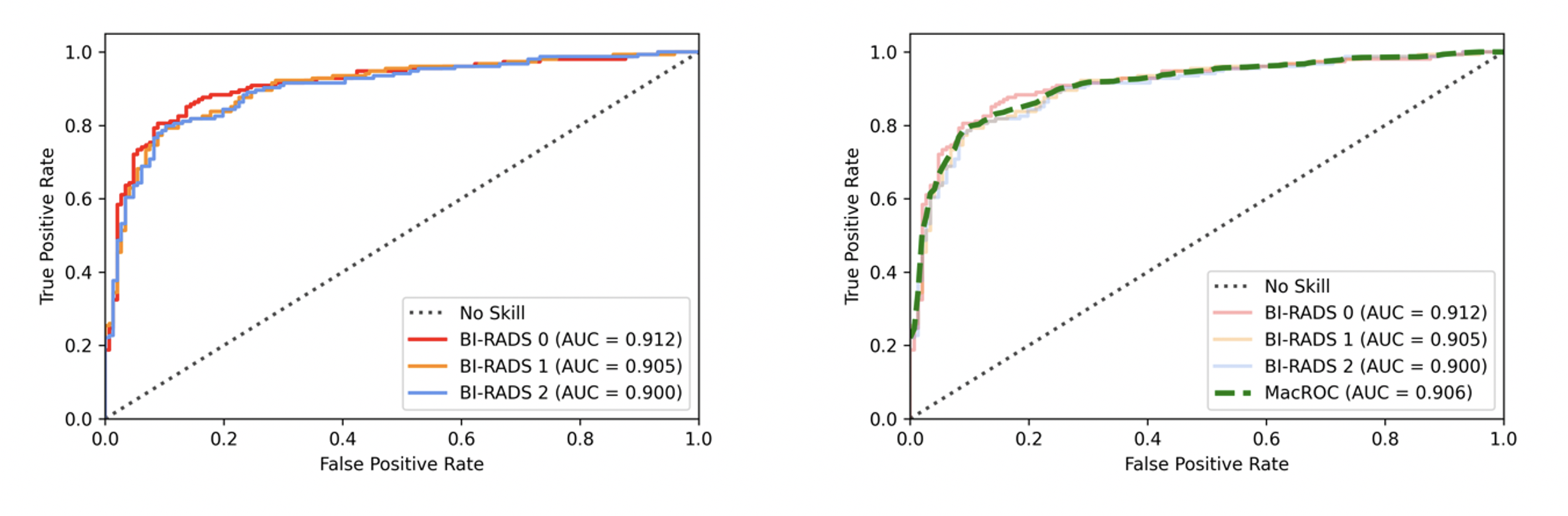}
  \caption{ROC curves computed with all test data for different BI-RADS classes (left) and average ROCs computed for all classes (right). AUC = area under the ROC curve.}
  \label{fig:fig1}
\end{figure}

\begin{figure}
  \centering
  \includegraphics[width=6cm]{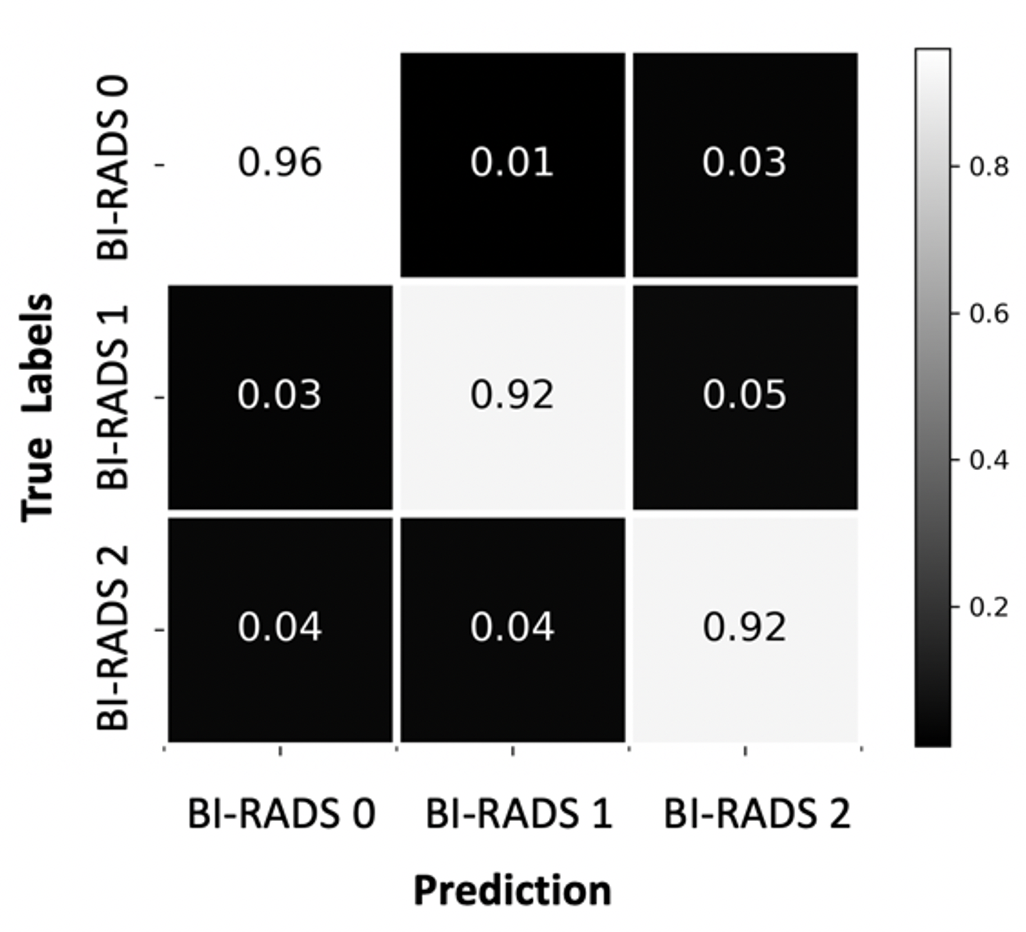}
  \caption{Confusion matrix (percent) on the testing dataset.}
  \label{fig:fig1}
\end{figure}

\subsection{Effect of Dataset Size}
Figure 4 illustrates the classification performance by changing the size of training dataset to the 100\%, 50\% and 25\% of the original dataset. As shown in the table (left), using the 100\% of dataset, the model achieves average AUC of 0.906 whereas decreasing the size of training set by 50\% and 25\% shows a drop in performance by average AUC=0.868, P=0.010 and AUC=0.817, P=0.009 respectively, which are significantly different from that with 100\% of the training set.

Overall performance for classification of different classes improves as the size of training set increases.

\begin{table}[]
\centering
\caption{The effect of changing the size of the dataset in terms of AUC metric }
\label{tab:my-table}
\begin{tabular}{r|c c c c c c c}
 \textbf{Fraction} & &\textbf{25\%} & & \textbf{50\%} & & \textbf{100\%} \\ \cline{1-7}
  \textbf{BI-RADS 0 vs. others} & & 0.834 & & 0.877 & & 0.912 \\ \cline{1-7}
  \textbf{BI-RADS 1 vs. others} & & 0.806 & & 0.868 & & 0.905 \\  \cline{1-7}
  \textbf{BI-RADS 2 vs. others} & & 0.812 & & 0.859 & & 0.900 \\ \cline{1-7}
  \textbf{Mac AUC} & & 0.817 & & 0.868 & & 0.906
\end{tabular}
\end{table}

\begin{figure}
  \centering
  \includegraphics[width=8cm]{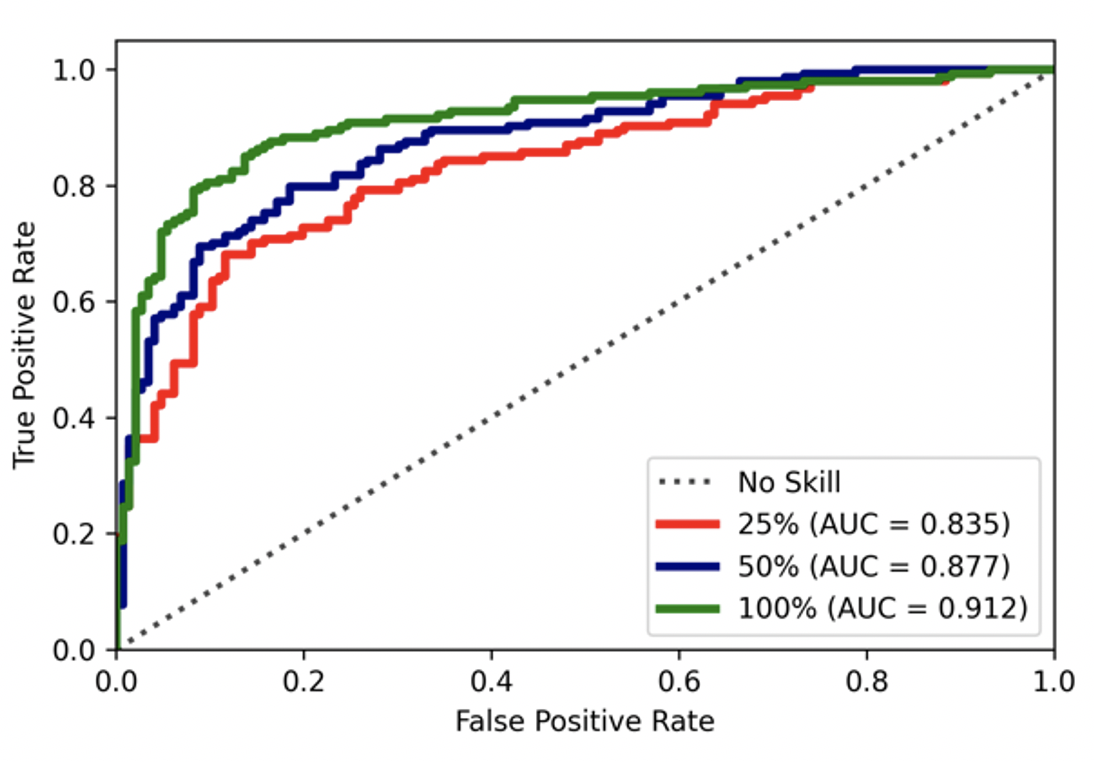}
  \caption{Comparison of ROC curves for BI-RADS 0 class. Increasing the size of dataset yields better results.}
  \label{fig:fig1}
\end{figure}

\subsection{Effect of Image Resolution}
As shown in Figure 5, at the highest resolution, the model achieves average AUC of 0.906 whereas down-sampling images by a factor of 1/2 yields a drop in performance with AUC=0.868 and P=0.011 which is significantly different from that for highest resolution images.

More aggressive down-sampling shows a larger drop in performance, with an average AUC of 0.817 and P=0.009 which is significantly different from that for highest resolution images. Further reduction in image resolution causes the size of the feature maps in later stages of network to become smaller than the convolutional kernels. In that case, we skipped the remaining layers before the global average pooling.

\begin{table}[]
\centering
\caption{TThe effect of changing the image resolution in terms of AUC metric.}
\label{tab:my-table}
\begin{tabular}{r|ccccccc}
 \textbf{Image Resolution} & &\textbf{$\times$1/4} & & \textbf{$\times$1/2} & & \textbf{$\times$1} \\ \cline{1-7}
  \textbf{BI-RADS 0 vs. others} & & 0.813 & & 0.870 & & 0.912 \\  \cline{1-7}
  \textbf{BI-RADS 1 vs. others} & & 0.787 & & 0.846 & & 0.905 \\  \cline{1-7}
  \textbf{BI-RADS 2 vs. others} & & 0.778 & & 0.840 & & 0.900 \\ \cline{1-7}
  \textbf{Mac AUC} & & 0.793 & & 0.852 & & 0.906
\end{tabular}
\end{table}

\begin{figure}
  \centering
  \includegraphics[width=8cm]{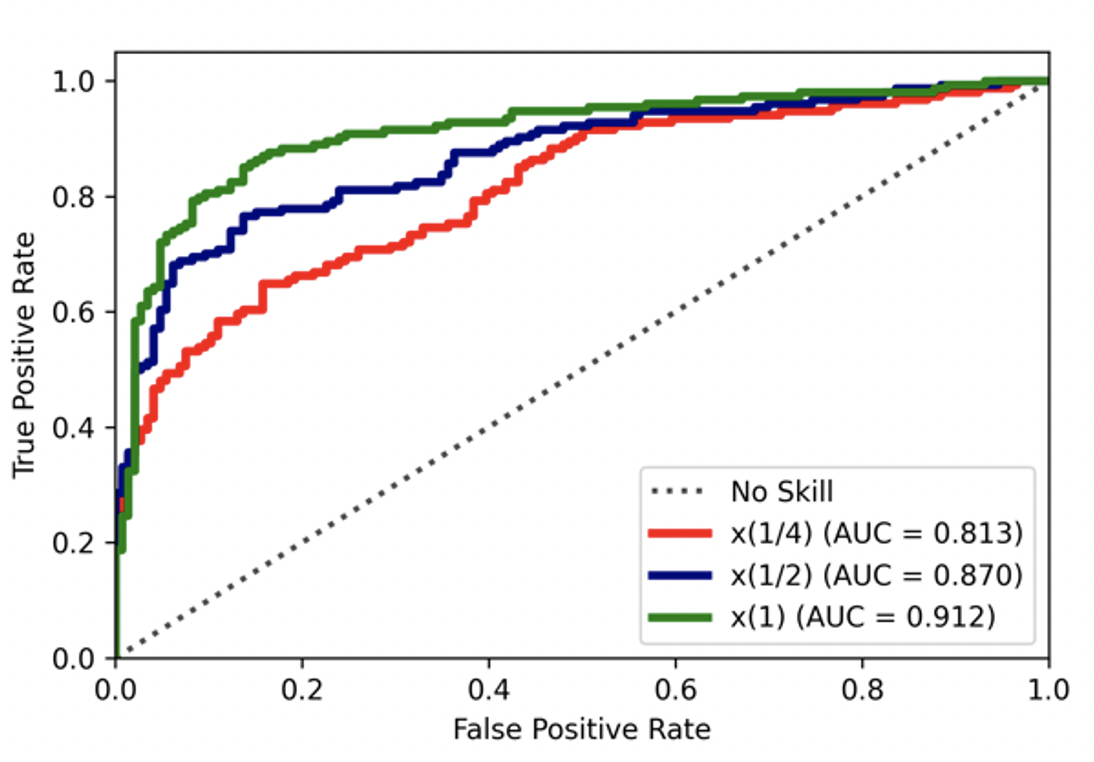}
  \caption{Comparison of ROC curves for BI-RADS 0 class with different image resolutions (right). Increasing the image resolution yields better results.}
  \label{fig:fig1}
\end{figure}

\subsection{Model Interpretation}

Figure 6 shows examples of saliency maps for a truly predicted BI-RADS 0 case indicating the regions that provide the highest contributions to the model’s predictions. 

\begin{figure}
  \centering
  \includegraphics[width=10cm]{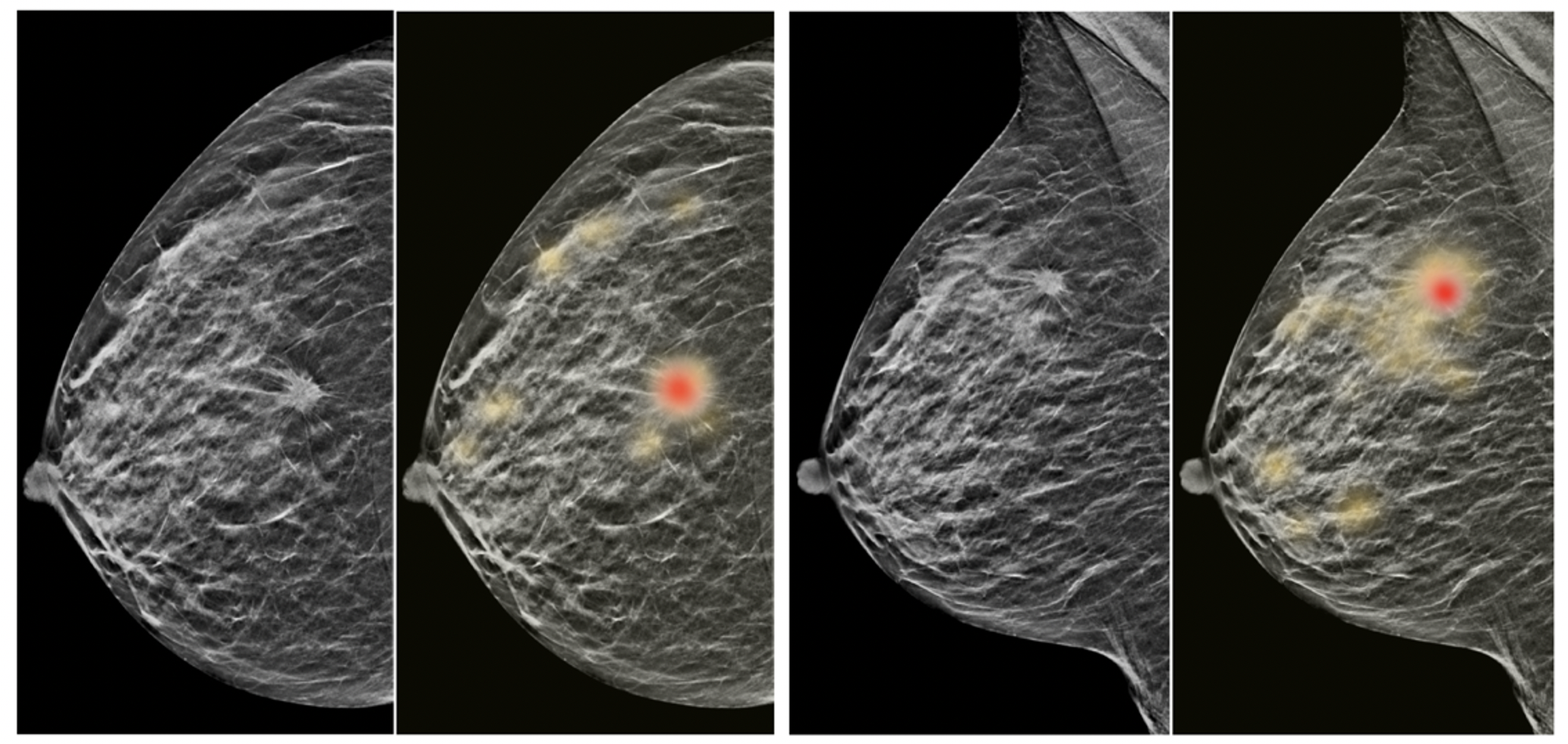}
  \caption{Example saliency maps for a truly predicted BI-RADS 0 case with abnormal findings as reported in the corresponding radiology reports.}
  \label{fig:fig1}
\end{figure}

\section{DISCUSSION}

In this study, we developed a novel end-to-end multi-view convolutional neural network model to classify high-resolution screening synthetic mammograms as normal, benign, or needing further workup. Our model achieved a high AUC for classification of BI-RADS 0 vs other classes, which was a key first step in building a classification model for screening breast digital tomosynthesis.  

We showed that reducing image resolution by unconditional down-sampling prior to training significantly reduced the model’s performance. At the same time, we studied the effect of dataset size on model performance. Similar to other deep learning applications, larger datasets result in higher performance. Our results demonstrate the importance of using dataset with at least tens of thousands of screening mammograms \cite{Dutta2018Evaluation}. 

Our study has limitations. Digital breast tomosynthesis data, includes reconstructed three-dimensional volumes alongside two-dimensional CC and MLO images. These volumes can frequently be used by mammographers to assist in disambiguating potentially abnormal findings initially observed in synthetic mammograms. In this study we did not include 3D volumes to reduce the computational and memory complexity of training 3D deep CNN. Incorporating these 3D volumes in model training process is a topic for future study. 

Radiologists typically have access to additional clinical data such as previous reports, imaging, and demographic information while reading the mammograms and use this additional information to support their decision. Our model was using only current images, which were labeled after processing the current reports. A more comprehensive study should incorporate all available information in a multi-modality manner to make the final decision for each case.

\section*{ACKNOWLEDGMENT}

Support for this work was provided by a gift from the Paustenbach Medical AI Research Fund.

\bibliographystyle{unsrt}  
\bibliography{screenet_paper_preprint_v1}  






\end{document}